\documentclass[12pt,a4paper]{article}
\pdfoutput=1
\usepackage{amsmath,amssymb,exscale}
\usepackage{amsfonts,amsthm}
\usepackage{graphicx, float, epsfig}
\usepackage{mathrsfs}
\usepackage{slashed}
\usepackage{textcomp}
\usepackage[T1]{fontenc}
\usepackage[utf8]{inputenc}
\usepackage[table]{xcolor}
\usepackage[lowtilde]{url}
\usepackage[colorlinks=true,urlcolor=blue,anchorcolor=blue,citecolor=blue,filecolor=blue,linkcolor=blue,menucolor=blue,pagecolor=blue,linktocpage=true]{hyperref}
\numberwithin{equation}{section}

\newcommand{\be}{\begin{equation}}
\newcommand{\ee}{\end{equation}}
\newcommand{\ba}{\begin{eqnarray}}
\newcommand{\ea}{\end{eqnarray}}
\newcommand{\bay}{\begin{array}{rcl}}
\newcommand{\eay}{\end{array}}

\newcommand{\bra}{\langle}
\newcommand{\ket}{\rangle}

\def\MP{M_{\rm Pl}}

\begin{document}

\title{Supersymmetric Preons and the Standard Model}

\author{Risto Raitio\footnote{E-mail: risto.raitio@gmail.com}\\	
02230 Espoo, Finland}

% Dec 13, 2017
\date{December 18, 2017} \maketitle

\abstract{\noindent
The experimental fact that standard model superpartners have not been observed compels one to consider an alternative implementation for supersymmetry. The basic supermultiplet proposed here consists of a photon and a charged spin 1/2 preon field, and their superpartners. These fields are shown to yield the standard model fermions, Higgs fields and gauge symmetries. Supersymmetry is defined for unbound preons only. Quantum group  SLq(2) representations are introduced to classify topologically scalars, preons, quarks and leptons.}
 
%04.70.-s   	    %Physics of black holes	 
%04.60.Bc 	    %Phenomenology of quantum gravity
%12.60.Rc	    %Composite models


\vskip 2.0cm
\noindent
Accepted for publication in Nuclear Physics B (25-Apr-18).

\newpage

%*********************
%\tableofcontents
\vskip 1.0cm

\section{Introduction}
\label{Intro}

What does it really mean that no standard model (SM) superpartners have been observed, at the LHC or elsewhere, but the SM is in good shape? Something seems to be undeniably ambiguous. Therefore an alternative point of view should be looked for. I believe the answer is that supersymmetry (SUSY) has to be implemented on a different level. I propose SUSY should be introduced one level below the SM, namely on the quark and lepton constituent, i.e. preon level. The beauty of SUSY is that a single fermion Dirac-Maxwell system plus superpartners is shown to produce the standard model fields: quarks, leptons, gauge fields and the Higgs. The preons themselves do not participate in strong and weak interactions \cite{Rai_00}, which are emergent phenomena in this model. 

The scheme presented in this article is based on a previous preon model of SM particles \cite{Rai_00}. I define the updated, supersymmetric model as follows
\begin{enumerate}
\item the elementary fields are members of a supermultiplet, 
\item the matter field is a light spin 1/2, charge 1/3 preon,
\item the gauge interaction is electromagnetism,
\item the quantum group SLq(2) is used to classify topologically scalar particles, preons, quarks and leptons \cite{Finkel_00, Finkel_0a}.
\end{enumerate}

Graviton and gravitino must also belong to the supermultiplet. Gravity is further supposed to organize the preons in bound state `bags'. These questions are, however, beyond the scope of this article.

The point of this article is that supersymmetry must be implemented in a specific way. Composite systems in physics often `hide' quantum numbers. Single hadrons are color neutral as measured in laboratory but quarks carry color quantum number inside hadrons. Following this method I propose that supersymmetry is valid only for free preons. Bound states of preons do not feel supersymmetry. 

This article is organized as follows. The supersymmetric preon model is described in section \ref{Preon}. Elements of supersymmetry are introduced in subsection \ref{susy}. After the apogee, in section \ref{Knot} quantum group structure for scalars, preons, quarks and leptons is described to indicate the differences and the unity of these objects. Finally, conclusions are given in section \ref{Concl}. The article is aimed to be self-contained. 

%*******************
\section{Supersymmetric Preon Model}
\label{Preon}

In case supersymmetry is familiar subsection \ref{susy} may be skipped by going to subsection \ref{preonmodel}.

%*************************** 
\subsection{Supersymmetry}
\label{susy}

An operator $Q$ which generates transformations between fermions and bosons is an anti-commuting spinor
\be
Q|\rm{boson}\ket = |\rm{fermion}\ket, ~~~ Q|\rm{fermion}\ket = |\rm{boson}\ket
\label{}
\ee

$Q$ and its hermitian conjugate $Q^{\dagger}$ carry spin 1/2. Therefore supersymmetry must be a space-time symmetry. The generators $Q$ and $Q^{\dagger}$ satisfy the following algebra
\be
	\begin{split}
	\{Q, Q^{\dagger}\} =&~ P^{\mu} \\
	\{Q, Q\} =&~ \{Q^{\dagger}, Q^{\dagger}\} = 0 \\
	[P^{\mu}, Q] =&~ [P^{\mu}, Q^{\dagger}] = 0
	\end{split}
\label{}
\ee
where $P^{\mu}$ is the four momentum generator of space-time translations.

Let us introduce the simplest possible supersymmetric model, the Wess-Zumino model \cite{Wess_Z}.\footnote{A fluent introduction to supersymmetry is \cite{Martin}.} It contains a single chiral supermultiplet: a massless, non-interacting left-handed two-component Weyl fermion $\psi$ and, as its superpartner, a complex scalar field $\phi$. The corresponding Lagrangian is
\be
\mathcal{L}_{WZ} = i\psi^{\dagger}\bar{\sigma}^{\mu}\partial_{\mu}\psi - \partial^{\mu}\phi^*\partial_{\mu}\phi
\label{WZlagrangian}
\ee
where $\bar{\sigma}^0 = \sigma^0, \bar{\sigma}^i = - \sigma^i, i=1, 2, 3$. The infinitesimal supersymmetry transformation of the scalar field changes it to a fermion
\be
\delta\phi = \epsilon\psi, ~~~\delta\phi^* = \epsilon^{\dagger}\psi^{\dagger}
\label{phitransform}
\ee
where $\epsilon^{\alpha}$ is an infinitesimal, anti-commuting, two-component Weyl fermion quantity. It turns out to be constant and has the dimension of [mass]$^{-1/2}$. For the fermion the transformation is
\be
\delta\psi_{\alpha} = -i\big(\sigma^{\mu}\epsilon^{\dagger}\big)_{\alpha}\partial_{\mu}\phi, ~~~ \delta\psi^{\dagger}_{\dot{\alpha}} = i\big(\epsilon\sigma^{\mu}\big)_{\dot{\alpha}}\partial_{\mu}\phi^*
\label{psitransform}
\ee

The supersymmetry algebra closes only on-shell, i.e. when the classical equations of motion are valid. One would like the symmetry to hold also quantum mechanically. For this purpose one introduces an auxiliary complex scalar field $F$ which has no kinetic term. It has dimension [mass]$^2$ and it gives a term $F^*F$ in the Lagrangian. The auxiliary field is useful in connection of spontaneous supersymmetry breaking but for the present purposes it cancels out.

In a general supersymmetric model there are one or more chiral supermultiplets with both gauge and non-gauge interactions. Let us construct a general model of masses and non-gauge interactions for particles contained in (\ref{WZlagrangian}). We want to find the most general renormalizable interactions for the Lagrangian (\ref{WZlagrangian}). Each term in the interaction Lagrangian $\mathcal{L}_{int}$ must have field content of mass dimension $\leq 4$. The only available terms are 
\be
\mathcal{L}_{int} = -\frac{1}{2}C^{ij}\psi_i\psi_j + C^i F_i + c^{ij}F_iF_j + c.c. -U
\label{intlagrangian}
\ee
where the $C^{ij}, C^{i}, c^{ij}$ and $U$ are polynomials in the scalar fields $\phi_i, \phi^{*i}$ with degrees 1, 2, 0 and 4, respectively.

Requiring (\ref{intlagrangian}) to be invariant under supersymmetry transformations leads to the condition that $U(\phi_{i}, \phi^{*i})$ must vanish. Likewise the coefficients $c^{ij}$ are zero, and we are left with $-1/2 C^{ij}\psi_i\psi_j + C^i F_i + c.c.$ only. Further it turns out that $W^{ij}$ cannot contain $\phi^{*i}$.  Therefore we have 
\be
C^{ij} = M^{ij} + d^{ijk}\phi_k
\label{}
\ee
whre $M^{ij}$ is a symmetric mass matrix for the fermion fields, and $d^{ijk}$ is a Yukawa coupling of a scalar $\phi_k$ and two fermions $\psi_i\psi_j$ that is totally symmetric under interchange of $i, j, k$. It is convenient to write 
\be
C^{ij} = \frac{\delta^2}{\delta\phi_i\delta\phi_j} W
\label{}
\ee
where the $W$
\be
W = \frac{1}{2}M^{ij}\phi_i\phi_j + \frac{1}{6}d^{ijk}\phi_i\phi_j\phi_k
\label{super-potential}
\ee
is called the superpotential. It is a holomorphic function of the complex scalar fields $\phi_i$. One can add a linear term $B^i\phi_i$ in the superpotential (\ref{super-potential}). The fermions and the bosons obey the same wave equation. Their masses are obtained from the same mass matrix with real non-negative eigenvalues: $(M^2)_i^{~j} = M^*_{ik}M^{kj}$.

The chiral Lagrangian density is
\be
\mathcal{L}_{chiral} = -\partial^{\mu}\phi^{*i}\partial_{\mu}\phi_i  + i\psi^{\dagger i}\bar{\sigma}^{\mu}\partial_{\mu}\psi_i - \frac{1}{2}\big(W^{ij}\psi_i\psi_j + W^*_{ij}\psi^{\dagger i}\psi^{\dagger j}\big) - W^i W^*_i
\label{chirallagrangian}
\ee

Details of interactions, as well as supersymmetry breaking, are beyond the scope of this article. I hope to return to these questions elsewhere. 

%*********************
\subsection{Preon Model}
\label{preonmodel}

The setup is the following. Quarks and leptons consist of three light spin 1/2, charge zero or 1/3 particles called preons, as indicated below (\ref{2-1}). At an energy of the order $10^{16\pm1}$ GeV quarks and leptons ionize into their constituents, preons. Below this dividing point, I assume that the standard model, with all its subtleties, is well behaving renormalizable theory. Above the ionization energy, supersymmetry enters the scene: it is defined for preons, which are now unbound. %This is reminiscent of quark color which operates only inside hadrons.

In the simplest supersymmetric preon model one has as the basic constituents the photon $\gamma$ and its neutral spin 1/2  superpartner, photino, denoted $\tilde{m}^0$. The second superpair is the charge 1/3, spin 1/2 preon $m^+$ and a complex scalar superpartner $\tilde{s}^+$. All fields $\gamma$, $\tilde{m}^0$, $m^+$ and $\tilde{s}^+$ have two degrees of freedom:  
\be
	\begin{split}
\scriptsize{\gamma = \left( \begin{array}{c} \rightarrow \\ \leftarrow \\ \end{array} \right)}~~ \rm{and}~~~
\scriptsize{\tilde{m}^0 = \left( \begin{array}{c} \uparrow \\ \downarrow \\ \end{array} \right)}, ~~
\scriptsize{m^+ = \left( \begin{array}{c} \uparrow \\ \downarrow \\ \end{array} \right)} ~~\rm{and}~~~
\scriptsize{\tilde{s}^+_{1,2}} 
	\end{split}
	\label{susyconstituents}
\ee
where the horizontal and vertical arrows refer to helicity and spin, respectively, and + and 0 refer to charge in units of 1/3 electron charge. $\tilde{s}^+_{1,2}$ indicates that there are two charged scalar fields $\tilde{s}^+_1 ~\rm{and}~ \tilde{s}^+_2$. The $m^+$, $\tilde{m}^0$ are assumed to have light mass, of the order of the first generation quark and lepton mass scale. The $\tilde{s}^+$ adds two physical scalar bosons, each consisting of three $\tilde{s}^+$ preons, with charge one. The R-parity for fields in (\ref{susyconstituents}) is simply $P_R = (-1)^{2(spin)}$. 

The standard model is constructed as follows. In section \ref{Knot} it will be established that matter originates from the $j=3/2$ and $j=1/2$ representations of  $SLq(2)$ for fermions. Originally this was postulated from a simple analysis of quark and lepton quantum numbers in \cite{Rai_00}. 

Starting from the known quark and lepton charges it is natural to assume the following charge quantization \{0, 1/3, 2/3, 1\} of which only the first (0) and the last (1) are physical charge states. To make these charges available with spin $1/2$ preons, quarks and leptons must be built as bound states of three preons. For same charge preons fermionic permutation antisymmetry factor $\epsilon_{ijk}$ must be included. These arguments lead to four bound states made of preons, which form the first generation quarks (q) and leptons (l) \cite{Rai_00} (dropping tildes)
\be
  \begin{split}
  u_k =&~ \epsilon_{ijk} m^+_i m^+_j m^0  \\
  \bar{d}_k =&~ \epsilon_{ijk} m^+ m^0_i m^0_j   \\
  e =&~ \epsilon_{ijk} m^-_i m^-_j m^-_k  \\
  \bar{\nu} =&~ \epsilon_{ijk} \bar{m}^0_i \bar{m}^0_j \bar{m}^0_k 
  \end{split} 
\label{2-1}
\ee

A useful feature in (\ref{2-1}) with two same charge preons, on lines 1 and 2, is that the construction provides a three-valued index for quarks on the left-hand side, to be identified as quark $SU(3)$ color. This is how color emerges in the preon model. It is the way quark color was originally discovered \cite{Greenberg}. The corresponding gauge boson states $q_i\bar{q_j}\rightarrow g_{ij}$ are in the adjoint representation $\bf{3}\otimes\bf{3^*}=\bf{8}\oplus\bf{1}$.\footnote{Introduction of color is done slightly differently in \cite{Finkel_0a}.} The weak $SU(2)$ left handed doublets can be read from the first two and last two lines in (\ref{2-1}): $l^-\bar{l^0}\rightarrow W^-$. The standard model (SM) gauge structure $SU(3)\times SU(2)$ is emergent in this sense from the present preon model. In the same way quark-lepton transitions between lines 1$\leftrightarrow$3 and 2$\leftrightarrow$4 in (\ref{2-1}) are possible gauge interactions. 

The horizontal lines in (\ref{2-1}) can be regarded as equivalence relations of different topological objects in the sense of section \ref{Knot}, and the vertical coordinate  in (\ref{2-1}) giving the gauge structure of the left-hand side fields.

The above gauge picture is supposed to hold up to the energy of about $10^{16 \pm 1}$ GeV. The electroweak interaction has the spontaneously broken symmetry phase below an energy of the order of 100 GeV and symmetric phase above it. The electromagnetic and weak forces take separate ways at higher energies ($100$ GeV$ \ll E \ll 10^{16}$ GeV), the latter restores its symmetry but melts away due to ionization of quarks and leptons into preons.This gives a cutoff for the theory. The electromagnetic interaction, in turn, stays strong towards Planck scale, $\MP \sim 1.22 \times 10^{19}$ GeV. Likewise, the quark color and leptoquark interactions suffer the same destiny as the weak force. One is left with the electromagnetic and gravitational forces only near Planck scale. 
 
The proton, neutron, electron and $\nu$ can be constructed of 12 preons and 12 anti-preons. The construction (\ref{2-1}) is matter-antimatter symmetric on preon level, which is desirable for early universe. The model makes it possible to create from vacuum a universe with only matter. Corresponding antiparticles may occur equally well, but the matter dominance case seems to have been made. Neutral dark matter is formed of preon-antipreon pairs more likely than ordinary matter when the temperature of the universe is lowered to a proper free mean path value between preon collisions.

The baryon number (B) is not conserved in this model: a proton may decay at Planck scale temperature by a preon rearrangement process into a positron and a pion. This is expected to be independent of the details of the preon interaction. Baryon number minus lepton number (B-L) is conserved.

%The preon model is approximately conformal at energies $\gg 10^{16}$ GeV. % Penrose, AdS/CFT

%***********************************
\section{Knot Theory of Fermions}
\label{Knot}

Early work on knots in physics goes back in time to 19th and 18th century \cite{Thom, Fadd_N}. More recently Finkelstein has proposed a model based on the quantum group  $SLq(2)$ \cite{Finkel_00, Finkel_0a}. The idea stems from the fact that Lie groups are degenerate forms of quantum groups \cite{Reshet_T_F}. Therefore it is of interest to study a physical theory by replacing its Lie group by the corresponding quantum group. Finkelstein introduced the global group $SLq(2)$ as an extension to the SM electroweak gauge group obtaining the group structure $SU(2) \times U(1) \times SLq(2)$.

Knots are objects in three dimensional space. Their projections onto two dimensional plane are considered here. Oriented knots can be characterized by three numbers as follows. Where two dimensional curves cross there is an overline and an underline at each point, a vertex. It has a crossing sign +1 or -1 depending on whether the overline direction is carried into the underline direction by a counterclockwise or clockwise rotation, respectively. The sum of all crossing signs is the writhe $w$ which is a topological invariant. The number of rotations of the tangent of the curve in going once around the knot is a second topological invariant and it is called the rotation $r$. An oriented knot can be labeled by the number of crossings $N$, the writhe $w$ and rotation $r$. The writhe and rotation are integers of opposite parity.

One can transform to quantum coordinates $(j, m, m')$. These indices label the irreducible representations of $D^j_{mm'}$ of the symmetry algebra of the knot, $SLq(2)$, by defining
\be
j = N/2, ~~ m = w/2, ~~ m' = (r + o)/2
\label{3-1}
\ee
This linear transformations makes half-integer representations possible. The knot constraints require $w$ and $r$ to be of opposite parity, therefore $o$ is an odd integer.

The knot or quantum algebra for a two dimensional representation $\scriptsize{M = \left( \begin{array}{cc} a & b \\ c & d \\ \end{array} \right)}$ is as follows
\be
  \begin{split}
  ab =&~ qba ~~~bd = qdb ~~~ad -qbc = 1 ~~~bc = cb\\
  ac =&~ qca ~~~cd = qdc ~~~da - q_1 cb = 1 ~~q_1 = 1/q
  \end{split}
\ee
with real valued q. 

Any knot $(N, w,r)$ may be labeled by $D^{N/2}_{w/2,(r+o)/2}(a, b, c, d)$. The following expression of the algebra is associated to a $(N, w, r)$ knot 
\be
	D^j_{mm'}(a. b, c, d) = \sum_{\substack{\delta(n_a+n_b, n_+) \\ \delta(n_c+n_d, n_-)}} A^j_{mm'}(q, n_a, n_c) \delta(n_a + n_b, n^{'}_+) a^{n_a}b^{n_b}c^{n_c}d^{n_d}
\label{3-2}
\ee
where $(j, m, m')$ is given by (\ref{3-1}), $n_{\pm} = j \pm m$, $n^{'}_{\pm} = j \pm m^{'}$ and $A^j_{mm'}(q, n_a, n_c)$ is given by
\be
A^j_{mm'}(q, n_a, n_c) = \bigg[\frac{\bra n^{'}_+\ket_1 \bra n^{'}_-\ket_1}{\bra n_+\ket_1 \bra n_-\ket_1} \bigg]^{1/2} \frac{\bra n_+\ket_1!}{\bra n_a\ket_1!\bra n_b\ket_1!} \frac{\bra n_-\ket_1!}{\bra n_c\ket_1!\bra n_d\ket_1!}
\label{3-3}
\ee
where $n_+ = n_a + n_b$, $n_- = n_c + n_d$, $\bra n\ket_q = \frac{q^{n-1}}{q-1}$ and $\bra\ket_1 = \bra\ket_{q_1}$. 

One assigns physical meaning to the $D^j_{mm'}$ in (\ref{3-2}) by interpreting the a, b, c, and d as creation operators for  spin 1/2 fermions. These are the four elements of the fundamental $j = 1/2$ representation $D^{1/2}_{mm'}$ as indicated in table 1.
\begin{center}
  \begin{tabular}{ c | c | c }
    \hline
    m & m' & preon \\ \hline
    ~1/2 & ~1/2 & a \\ \hline
    ~1/2 & -1/2 & b \\ \hline
    -1/2 & ~1/2 & c \\ \hline
    -1/2 & -1/2 & d \\
    \hline
  \end{tabular} \\~\\
\bf{Table 1. }  \\
\textmd{The j=1/2 representation of SLq(2) giving the preon states.}
\end{center}
Using the terminology of section \ref{Preon}, let us call the fermions $j=1/2$ preons and $j=3/2$ quarks or leptons. I use in the tables 1. and 2. the fermion names of \cite{Finkel_00}. The preon dictionary from the notation of \cite{Rai_00} to the notation of \cite{Finkel_00} is the following:
\be
  \begin{split}
  m^+ \mapsto  a, ~~ m^0 \mapsto  c \\
  m^- \mapsto d, ~~ \bar{m^0} \mapsto  b
  \end{split}
\label{}
\ee

The standard model particles are the following $D^{3/2}_{mm'}$ representations 
\begin{center}
  \begin{tabular}{ l | c | c | c }
    \hline
    ~~m & m' & particle & preons \\ \hline
    ~3/2 & 3/2 & electron & aaa \\ \hline
    ~3/2 & 3/2 & neutrino & ccc \\ \hline
    ~3/2 & -1/2 & d-quark & abb \\ \hline
    -3/2 & -1/2 & u-quark & cdd\\
    \hline
  \end{tabular} \\~\\
\bf{Table 2. } \\
\textmd{The j=3/2 representation of SLq(2) giving the the first generation quark and lepton states.}
\end{center}

All details of the $SLq(2)$ extended standard model are discussed in \cite{Finkel_00}, including the composite gauge and Higgs bosons and a candidate for dark matter. I do not, however, see much advantage for introducing composite gauge bosons in the model (gauge invariance is a local property). Therefore the model of \cite{Rai_00} and the knot algebra model of \cite{Finkel_00} are equivalent in the fermion sector.

Preon binding into bound states is not completely clear. The trefoil field structure may be regarded as a trefoil flux tube carrying energy, momentum and charge so that all three are concentrated at the three crossings. Then one can regard these three concentrations at the three crossings as actually defining the three preons, without assuming their existence as independent degrees of freedom. 

The most elementary configuration is a simple loop having $j=0$. These are scalar fields. Some pairs of these loops with opposite rotation may be brought together, e.g. in the early universe, by gravitational attraction making two opposing $j = 1/2$ twisted loops.

In summary, knots having odd number of crossings are fermions and knots with even number of crossings are correspondingly bosons. Instead of considering spin 1 bound states of six preons I assumed in section \ref{Preon} that the SM gauge bosons are genuine point like gauge fields. The leptons and quarks are simple quantum knots, the quantum trefoils, with three crossings and $j = 3/2$. At each crossing there is a preon. The preons are twisted loops with one crossing and $j = 1/2$. The $j = 0$ states are simple neutral loops with zero crossings (see figures in \cite{Finkel_00}). 

%***********************************
\section{Conclusions} 
\label{Concl}

The present supersymmetric preon model is an economic Dirac-Maxwell system. Its low energy limit is the standard model. The standard model can be constructed with the superpartners being included within the resulting model. 

Compared to the early version of the preon model the new fields are the two scalar spreons $\tilde{s}$ having charge 1/3. Consequently, there must be two three spreon scalar boson bound states with charge one in nature. Preon-antipreon and spreon-antispreon pairs make two neutral observable scalar boson. It is tempting to associate both the charged bound states and the neutral  preon-antipreon pairs with the Higgs bosons \cite{Branco_al}. Even single charge 1/3 scalar bosons may exist, which would be a clear signal for the model.

The SLq(2) group provides a solid topological basis for the fermion sector of the present model. In \cite{Finkel_0a} it was shown that the SLq(2) preon model agrees with the Harari-Shupe (H-S) rishon model \cite{Harari, Shupe}.\footnote{\scriptsize The basic idea of the present model was originally conceived during the week of the $\psi$ discovery in November 1974  at SLAC. I proposed that the c-quark is a gravitationally excited u-quark, both consisting of three spin 1/2 and charge \{0, 1/3\} heavy black hole constituents.   This idea met resistance. Therefore the model was not written down until years later.} Unlike the said preon/rishon authors, I do not think that (i) SM gauge bosons should be treated as bound states of several preons, or (ii) hypercolor is realistic for preon interactions. In any case, a mechanism for preon binding into quarks and leptons is yet to be developed. There are tentative models where the electron, or preon, is considered a spinning Kerr black hole with a superconducting `bag' vacuum source \cite{Burinskii}.

It is hoped that the present preon model provides a new avenue towards better understanding of the roles of all four interactions, and gravity in particular. In the present scenario, gravity and electromagnetism are the genuine interactions of early cosmology. The weak and strong interactions are emergent from the basic fermion structure of the model in (\ref{2-1}). Traditional advantages of supersymmetry (like top-stop cancellation, gauge coupling unification, etc.) loose their meaning in their old form. Instead, a wholly new phenomenology will have to be calculated, with the hope of including in it local supersymmetry. 

%***********************************
%\section*{Acknowledgement}

%\newpage

\end{document}